\documentclass[12pt, a4paper]{article}

\usepackage{graphicx}

\usepackage{stmaryrd}
\usepackage{overcite}
\usepackage{verbatim}
\usepackage{achemso}
\usepackage{verbatim}

\usepackage{multirow}
\usepackage{amsmath}
\usepackage{amsfonts}
\usepackage{amsgen}
\usepackage{longtable}
\usepackage{lscape}
\usepackage{threeparttable}
\usepackage{supertabular}
\usepackage{multicol}
\usepackage{multirow}
\usepackage{setspace}
\usepackage{rotating}
\usepackage{amssymb}
\usepackage{layout}
\usepackage{wrapfig}
\usepackage{subfig}

\begin{document}

\begin{center}

{\bf\large{One Electron Oxygen Reduction in Room Temperature Ionic Liquids: A Comparative Study of Butler-Volmer \\and Symmetric Marcus-Hush Theories Using Microdisc Electrodes}}

\hspace{2cm}

{\bf\large Eden E. L. Tanner, Linhongjia Xiong, Edward O. Barnes and Richard G. Compton*}

*Corresponding author\\
Email:~richard.compton@chem.ox.ac.uk\\
Tel:~+44~(0)~1865~275957.\\

Department~of~Chemistry, Physical~and~Theoretical~Chemistry~Laboratory, Oxford~University,
South~Parks~Road, Oxford, OX1~3QZ, United~Kingdom.
\vspace{1cm}

NOTICE: this is the authors version of a work that was accepted for publication in \emph{The Journal of Electroanalytical Chemistry}. Changes resulting from the publishing process, such as peer review, editing, corrections, structural formatting, and other quality control mechanisms may not be reflected in this document. Changes may have been made to this work since it was submitted for publication. A definitive version was subsequently published in \emph{The Journal of Electroanalytical Chemistry,}doi:10.1016/j.jelechem.2014.05.022

\end{center}

\clearpage

\section*{Abstract}

The voltammetry for the reduction of oxygen at a microdisc electrode is reported in two room temperature ionic liquids: 1-butyl-1-methylpyrrolidinium \emph{bis}(trifluoromethylsulfonyl)imide \\ ([Bmpyrr][NTf$_2$]) and trihexyltetradecylphosphonium \emph{bis}(trifluoromethylsulfonyl)imide \\([P$_{14,6,6,6}$][NTf$_2$]) at 298 K. Simulated voltammograms using Butler-Volmer theory and symmetric Marcus-Hush (SMH) theory were compared with experimental data. Butler-Volmer theory consistently provided experimental parameters with a higher level of certainty than SMH theory. A value of solvent reorganisation energy for oxygen reduction in ionic liquids was inferred for the first time as 0.4-0.5 eV, which is attributable to inner-sphere reorganisation with a negligible contribution from solvent reorganisation. The developed Butler-Volmer and symmetric Marcus-Hush programs are also used to theoretically study the possibility of kinetically limited steady state currents, and to establish an approximate equivalence relationship between microdisc electrodes and spherical electrodes resting on a surface for steady state voltammetry for both Butler-Volmer and symmetric Marcus-Hush theory.

\section{Introduction}

Current understanding of interfacial electron transfer (ET) kinetics is dominated by two theories, Butler-Volmer\cite{Butler1924, Erdey-Gruz1930} and Marcus-Hush\cite{Marcus1985, Hush1999}, which both describe a one-electron transfer:
\begin{equation}
\mathrm{A} + \mathrm{e^-} \rightleftharpoons \mathrm{B}
\end{equation}
The first, Butler-Volmer theory (BV, Equations \ref{BV RED} and \ref{BV OX}), is well-reported and reliable for parameterisation\cite{Laborda2012}, but is phenomenological, and therefore is unable to provide any detailed physical insights of a predictive nature. BV theory relates the electrode kinetics to the heterogeneous rate constants ($k^0$), the formal potential ($E_\mathrm{f}^0$), and the transfer coefficient, $\alpha$, which empirically describes how product or reactant-like the transition state is\cite{Compton2010}.
\begin{eqnarray}
k_\mathrm{red} = k^0\text{exp}\left[-\alpha\frac{F\left(E - E_\mathrm{f}^\minuso\right)}{RT}\right] \label{BV RED}\\
k_\mathrm{ox} = k^0\text{exp}\left[\left(1-\alpha\right)\frac{F\left(E - E_\mathrm{f}^\minuso\right)}{RT}\right] \label{BV OX}
\end{eqnarray}

An alternative theory, symmetric Marcus-Hush (SMH), offers greater microscopic physical insight into the system under study. MH theory relates the heterogeneous rate constant to the reaction Gibbs energy, $\Delta G^\minuso$, and reorganisation energy, $\lambda$:
\begin{eqnarray}
\Delta G^{\ddagger} = \frac{\lambda}{4}\left(1 + \frac{\Delta G^\minuso}{\lambda}\right)^2\\
k^0 = A\text{ exp}\left[\frac{-\Delta G^{\ddagger}}{RT}\right]
\end{eqnarray}
As is shown in Equation \ref{INNER PLUS OUTER}, the total reorganisation energy is the sum of contributions from inner sphere bond reorganisation energy ($\lambda_\mathrm{i}$) and solvent reorganisation energy ($\lambda_\mathrm{o}$). These can be estimated separately, with the outer-sphere solvent reorganisation energy from the Born-Marcus solvation equation\cite{Bron1920}, and inner-sphere contributions from computational calculations (for example, DFT etc\cite{Hartnig2002}) of changes in bond lengths and angles of the molecule.
\begin{equation}
\lambda = \lambda_\mathrm{i} + \lambda_\mathrm{o} \label{INNER PLUS OUTER}
\end{equation}

These two models have been used to analyse a range of electrochemical processes by cyclic voltammetry in aqueous\cite{Henstridge2011, Wang2012a} and non-aqueous\cite{Suwatchara2012, Suwatchara2011, Laborda2011} solvents. A comprehensive review by Henstridge \emph{et al}\cite{Henstridge2012f} concluded that SMH theory struggles to adequately describe the voltammetric response of solution-phase redox couples whose transfer coefficients deviate significantly from 0.5. A fourth parameter, $\gamma$, was introduced, which accounts for any differences between the inner-shell force constant of the reactant and the product, to create asymmetric Marcus-Hush theory (AMH). This refinement was explored in aqueous\cite{Laborda2012, Henstridge2012d} and non aqueous\cite{Laborda2013a, Suwatchara2012a, Suwatchara2012b} solvents and surface bound systems\cite{Henstridge2012g} by Laborda \emph{et al}, and it was concluded that AMH was able to parameterise the experimental voltammetry at least as well as BV theory, and more accurately than SMH theory.

The electroreduction of oxygen has been extensively studied in a variety of solvents\cite{Buzzeo2003, Katayama2005, Zhang2014, Wadhawan2003, AlNashef2001, Hayyan2011}. The product of the initial one-electron reduction, superoxide, is a highly nucleophilic ion that initiates further reactions with a proton source, if one is present in solution. However, in absence of a proton source, the superoxide ion is stable, and the reaction proceeds as a simple one electron transfer:
\begin{equation}
\mathrm{O}_2 + \mathrm{e^-} \rightleftharpoons \mathrm{O}_2^{\bullet -}\label{oxygenreduction}
\end{equation}
ET kinetics of irreversible oxygen reduction has been studied in-depth in aqueous solvents\cite{Hartnig2002, Gotti2013} and values for the total reorganisation energy for the reduction of oxygen to superoxide have been estimated to lie between 1.26-4.51 eV\cite{Eberson1987}. This has been attributed to rearrangement of the polar water molecules upon formation of the superoxide ions.

Room temperature ionic liquids (RTILs) are molten salts with melting points below room temperature\cite{Wasserscheid2000}, and normally consist of a bulky, asymmetric organic cation, and an inorganic anion\cite{Hussey1988}. Ionic liquids are interesting solvents in which to examine the reversible one-electron reduction of oxygen, because the superoxide has been known to be stable, leading to quasi-reversible voltammetry in a range of ILs.\cite{Buzzeo2003, Huang2009, Evans2004,Zhang2004,AlNashef2001}.

Lynden-Bell has conducted a thorough study into the applicability of Marcus theory to RTIL solvents\cite{Lynden-Bell2007, Lynden-Bell2007a}. It was suggested \emph{via} simulation that the outer sphere reorganisation energy for redox processes in RTILs will be similar to that for the same process in a non ionic polar solvent. In a polar solvent, the molecules will re-orient themselves in response to the addition of charge, and in an ionic liquid the ions will translate themselves. It was predicted that these two processes produce similar reorganisation energies.

The mechanism and ET kinetics have been elucidated for oxygen reduction to hydrogen peroxide in protic ionic liquids,\cite{Khan2013} and in hydrophobic RTILs in the presence of water.\cite{Rollins2009} As yet, the ET kinetics of the one electron reduction of oxygen to superoxide in dry ionic solvents has not been studied, and no values for reorganisation energy for this process have been established.

\begin{figure}[h]
\begin{center}
\includegraphics[width = 0.9\textwidth]{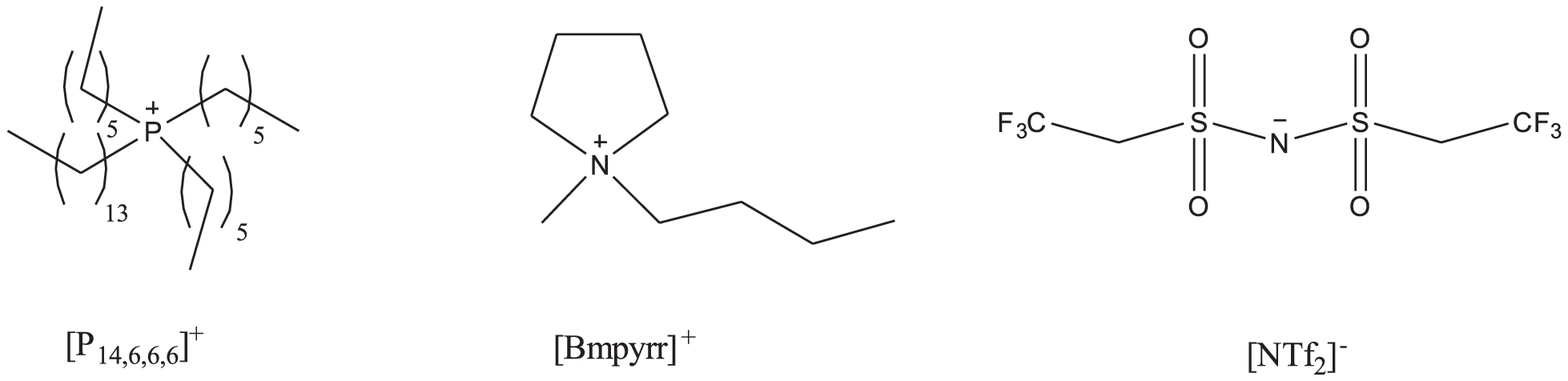}
\caption{RTILs used in this study; trihexyltetradecylphosphonium \emph{bis}(trifluoromethylsulfonyl)imide [P$_\mathrm{14,6,6,6}$][NTf$_2$] and 1-butyl-1-methylpyrrolidinium \emph{bis}(trifluoromethylsulfonyl)imide [Bmpyrr][NTf$_2$].} \label{IL STRUCTURES}
\end{center}
\end{figure}

This study is the first to examine SMH theory of one electron oxygen reduction in ILs at microdisc electrodes.  The structures of the ions in the ILs used in this study are shown in Fig. \ref{IL STRUCTURES}. The present work aims to develop the theory of SMH voltammetry at microdisc electrodes and experimentally determine a value for solvent reorganisation, $\lambda$, for oxygen reduction in two ionic liquids. The present work also aims to compare the ability of SMH theory and BV theory to simulate experimental reduction of oxygen and quantify a goodness of fit for each. This will allow conclusions to be reached about the relative ability of each theory to parameterise experimental data.

\section{Theory} \label{THEORY}

We consider a one electron reduction of an electroactive species:
\begin{equation}
\mathrm{A} + \mathrm{e^-} \rightleftharpoons \mathrm{B}
\end{equation}
at a microdisc electrode, using both Butler-Volmer theory and symmetric Marcus-Hush theory to describe the electron transfer kinetics. Cyclic voltammetry is simulated using both models, and both are subsequently compared to experimental data (see section \ref{RESULTS AND DISCUSSION}).

This study utilises microdisc electrodes, which offer several advantages over macroelectrodes. Firstly, as the capacitance of the double layer is proportional to the size of the electrode, a consequence of the reduced size of the electrode will be naturally lower capacitance. Secondly, the small currents passed reduce the ohmic drop and allow a two-electrode system to be employed\cite{Compton2010}. Lastly, the smaller area enables a very small volume of ionic liquid to be studied (in a T-cell system, as described previously\cite{Rodgers2008b}), which is desirable for logistical reasons (expense, ability to dry the aliquot of IL more quickly).

\begin{figure}[h]
\begin{center}
\includegraphics[width = 0.9\textwidth]{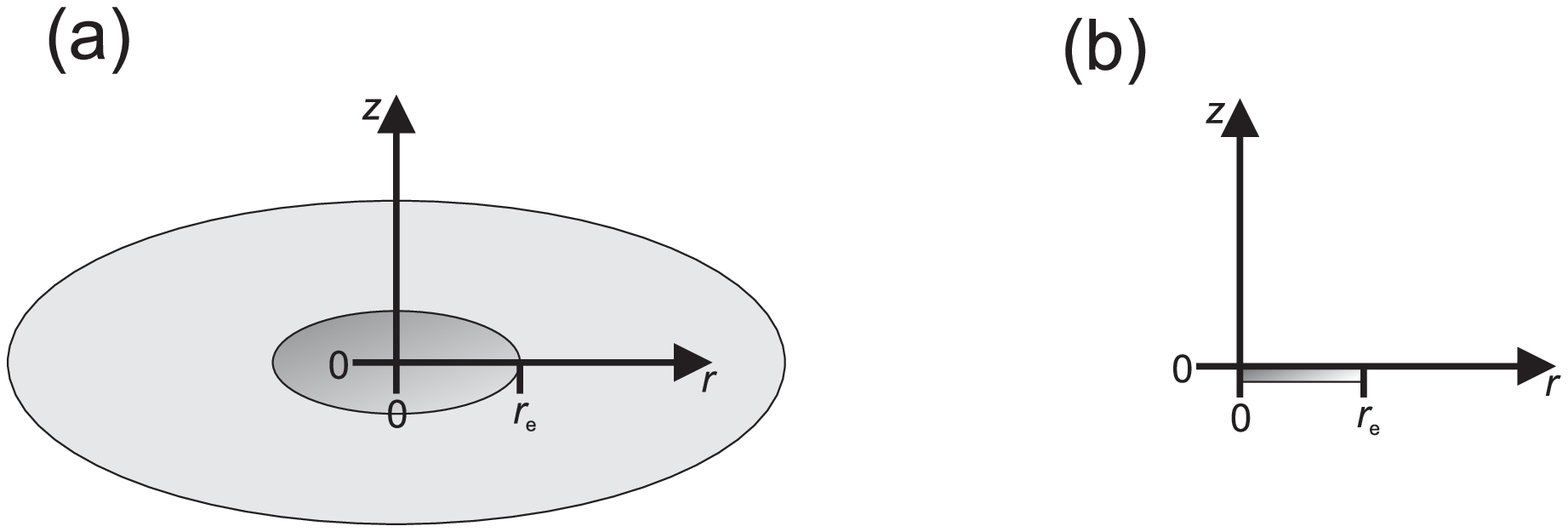}
\caption{(a) Schematic diagram of the microdisc electrode inlaid in a surrounding insulating material, defining the spatial coordinate system used in simulations. (b) The two dimensional simulation space used in this study.} \label{ELECTRODE SCHEMATIC}
\end{center}
\end{figure}

Cylindrical coordinates are employed to model the two dimensional microdisc electrode and simulate cyclic voltammetry. Fig. \ref{ELECTRODE SCHEMATIC} shows a schematic diagram of the microelectrode (a) and the two dimensional cylindrical space used in our model (b). Also defined are the spatial coordinates $r$ and $z$, as well as the electrode radius $r_\mathrm{e}$. In cylindrical space, Fick's second law for diffusional mass transport is:
\begin{equation}
\frac{\partial{c}_\mathrm{i}}{\partial{t}} = D_\mathrm{i}\left(\frac{\partial^2c_\mathrm{i}}{\partial{r^2}} + \frac{\partial^2c_\mathrm{i}}{\partial{z^2}} + \frac{1}{r}\frac{\partial{c}_\mathrm{i}}{\partial{r}}\right)
\end{equation}
All symbols are defined in Table \ref{DIMENSIONAL}. The mass transport is considered to take place under full support (as is usually assumed when ionic liquids are used as solvents), and therefore electrical migration can be neglected. Further, the absence of mechanical stirring and any significant temperature gradients eliminate the presence of forced and natural convection respectively, allowing a diffusion only model to be used.

The electron transfer reaction is only considered to occur at the electrode surface ($r \leq r_\mathrm{e}$, $z = 0$). The rate of this reaction is dependent on the potential applied to the electrode, $E$ (V). In cyclic voltammetry, the potential is given by the equation:
\begin{equation}
E = |E_\mathrm{s} - E_\mathrm{v} - \nu t| + E_\mathrm{v}
\end{equation}
where $E_\mathrm{s}$ and $E_\mathrm{v}$ are respectively the initial applied potential and the vertex potential (V), $\nu$ is the scan rate (V s$^{-1}$) and $t$ is the time since the start of the experiment (s). The flux of species A across the electrode due to electrolysis is then described as:
\begin{equation}
D_\mathrm{A}\frac{\partial{c_\mathrm{A}}}{\partial{z}} = k_\mathrm{red}c_\mathrm{A}^0 - k_\mathrm{ox}c_\mathrm{B}^0
\end{equation}
where the rate constants $k_\mathrm{red}$ and $k_\mathrm{ox}$ are functions of the applied potential $E$, and $c_\mathrm{A}^0$ and $c_\mathrm{B}^0$ are the concentrations, at the electrode surface, of species A and B respectively. The mathematical forms of $k_\mathrm{red}$ and $k_\mathrm{ox}$ depend on whether Butler-Volmer or Marcus-Hush kinetics are employed, as detailed below.

\subsection{Butler-Volmer Kinetics}

Within the Butler-Volmer formalism of electron transfer kinetics, $k_\mathrm{red}$ and $k_\mathrm{ox}$ are considered to be exponentially dependent on the applied potential:\cite{Butler1924}
\begin{eqnarray}
k_\mathrm{red} = k^0\mathrm{exp}\left[-\alpha\theta\right] \\
k_\mathrm{ox} = k^0\mathrm{exp}\left[\left(1-\alpha\right)\theta\right]
\end{eqnarray}
where $k^0$ is the standard heterogeneous electrochemical rate constant (m s$^{-1}$), $\alpha$ is the transfer coefficient and $\theta$ is a dimensionless potential, defined as:
\begin{equation}
\theta = \frac{F}{RT}\left(E - E^\minuso_\mathrm{f}\right)
\end{equation}
where $E^\minuso_\mathrm{f}$ is the formal potential of the A/B couple and $F$, $R$, and $T$ have their usual meanings.

The transfer coefficient $\alpha$, as described in the introduction, has a value of between zero and one, and is considered to reflect the structure of the transition state of the A/B redox couple\cite{Compton2010}.

\subsection{Marcus-Hush Kinetics}

In (symmetric) Marcus-Hush theory\cite{Marcus1956, Marcus1956a, Marcus1957, Marcus1957a, Marcus1964, Hush1958, Chidsey1991, Feldberg2010, Henstridge2012c}, $k_\mathrm{red}$ and $k_\mathrm{ox}$ are functions of the applied electrode potential and the reorganisation energy, $\lambda$. Mathematically:
\begin{eqnarray}
k_\mathrm{red} = k^0\frac{S_\mathrm{red}\left(\theta, \lambda\right)}{S_\mathrm{red}\left(0, \lambda\right)} \\
k_\mathrm{ox} = k^0\frac{S_\mathrm{ox}\left(\theta, \lambda\right)}{S_\mathrm{ox}\left(0, \lambda\right)}
\end{eqnarray}
$S_\mathrm{red/ox}\left(\theta, \lambda\right)$ is given by the following integral:
\begin{equation}
S_\mathrm{red/ox}\left(\theta, \lambda\right) = \int^\infty_{-\infty}{\frac{\text{exp}\left[-\Delta G^\ddagger_\mathrm{red/ox}/RT\right]}{1 + \text{exp}\left[\mp x\right]}}\text{d}x
\end{equation}
$\Delta G^\ddagger_\mathrm{red/ox}$ is the activation energy of reduction/oxidation, and is given by:
\begin{equation}
\frac{\Delta G^\ddagger_\mathrm{red/ox}}{RT} = \frac{\Lambda}{4}\left(1 \pm \frac{\theta + x}{\Lambda}\right)
\end{equation}
where $\Lambda$ is the dimensionless reorganisation energy, given by:
\begin{equation}
\Lambda = \frac{F}{RT}\lambda
\end{equation}
$x$ is an integration variable, defined as:
\begin{equation}
x = \frac{F}{RT}\left(\epsilon - E\right)
\end{equation}
This accounts for the continuum of energy levels in the electrode of energy $\epsilon$.

\subsection{Simulation Procedure}

To simulate cyclic voltammetry, the mass transport equation and electron transfer kinetic equations are normalised to remove scaling factors from the simulation. Substituting dimensionless parameters as defined in Table \ref{DIMENSIONLESS} into the mass transport equation, we obtain:
\begin{equation}
\frac{\partial{C}_\mathrm{i}}{\partial{\tau}} = D^{'}_\mathrm{i}\left(\frac{\partial^2C_\mathrm{i}}{\partial{R^2}} + \frac{\partial^2C_\mathrm{i}}{\partial{Z^2}} + \frac{1}{R}\frac{\partial{C}_\mathrm{i}}{\partial{R}}\right)
\end{equation}
The boundary condition at the electrode surface ($R \leq 1$, $Z = 0$) becomes:
\begin{equation}
\frac{\partial{C_\mathrm{A}}}{\partial{Z}} = K_\mathrm{red}C_\mathrm{A}^0 - K_\mathrm{ox}C_\mathrm{B}^0
\end{equation}
where
\begin{eqnarray}
K_\mathrm{red} = K^0\text{exp}\left[-\alpha\theta\right] \\
K_\mathrm{ox} = K^0\text{exp}\left[\left(1-\alpha\right)\theta\right]
\end{eqnarray}
for Butler-Volmer kinetics, and
\begin{eqnarray}
K_\mathrm{red} = K^0\frac{S_\mathrm{red}\left(\theta, \Lambda\right)}{S_\mathrm{red}\left(0, \Lambda\right)} \\
K_\mathrm{ox} = K^0\frac{S_\mathrm{ox}\left(\theta, \Lambda\right)}{S_\mathrm{ox}\left(0, \Lambda\right)}
\end{eqnarray}
for Marcus Hush kinetics. In both cases the dimensionless standard heterogeneous electrochemical rate constant is defined as:
\begin{equation}
K^0 = \frac{k^0r_\mathrm{e}}{D_\mathrm{A}}
\end{equation}

At the electrode surface, equal and opposite flux acts as the boundary condition for species B:
\begin{equation}
D^{'}_\mathrm{B}\frac{\partial{C_\mathrm{B}}}{\partial{Z}} = -\frac{\partial{C_\mathrm{A}}}{\partial{Z}}
\end{equation}
The outer boundaries of the simulation space are set at:
\begin{eqnarray}
R_\mathrm{max} = 1 + 6\sqrt{D^{'}_\mathrm{max}\tau_\mathrm{max}} \\
Z_\mathrm{max} = 6\sqrt{D^{'}_\mathrm{max}\tau_\mathrm{max}}
\end{eqnarray}
where $D^{'}_\mathrm{max}$ and $\tau_\mathrm{max}$ are, respectively, the largest dimensionless diffusion coefficient in the system and the total dimensionless time of the experiment. These boundaries have been shown to be sufficiently distant from the electrode to be well outside the depletion zone\cite{Gavaghan1998b, Gavaghan1998a, Gavaghan1998}.

At these boundaries, as well as at the symmetry boundary at $R = 0$, all $Z$ and at the insulating surface ($R > 1$, $Z = 0$), a zero flux condition is imposed:
\begin{equation}
\frac{\partial{C_\mathrm{i}}}{\partial{N}} = 0
\end{equation}
where $N$ is one of the spatial coordinates $R$ or $Z$ as appropriate.

Initial conditions are $C_\mathrm{A} = 1$ and $C_\mathrm{B} = 0$ throughout the simulation space. At $\tau = 0$ the experiment starts and the electrode surface boundary conditions are imposed. The total dimensionless flux $J$ at the electrode is calculated as:
\begin{equation}
J = \int^1_0{R\left(\frac{\partial{C_\mathrm{A}}}{\partial{Z}}\right)_{Z=0}}\text{d}R
\end{equation}
This dimensionless flux is transformed into a real current $I$ by:
\begin{equation}
I = 2\pi FD_\mathrm{A}c_\mathrm{A}^{*}r_\mathrm{e}J
\end{equation}

\subsection{Homogeneous kinetics}

In more complex reaction mechanisms, the single electron transfer thus far simulated may be followed by homogeneous kinetics. For example, the reduction of oxygen in the ionic liquid [P$_{14,6,6,6}$][NTf$_2$] (see Section \ref{P14666NTf2 RESULTS}) follows the following mechanism:
\begin{eqnarray}
\mathrm{O_2} + \mathrm{e^-} & \rightleftharpoons & \mathrm{O_2^{\bullet -}}\\
\mathrm{O_2^{\bullet -}} & \rightarrow & \mathrm{X}\\
\mathrm{X} & \rightleftharpoons & \mathrm{Y} + \mathrm{e^-}
\end{eqnarray}
where the homogeneous reaction involves proton abstraction by the superoxide ion from the phosphonium cation in the ionic liquid. This irreversible, pseudo first order reaction forms a product which can then be oxidised, at less negative potentials than the initial electron transfer to oxygen. The homogeneous step is considered to have a fast rate constant.

This being the case, if only the forward, reductive, wave of a cyclic voltammogram is measured and simulated, species X and Y may be neglected and only the first 2 steps in the above mechanism considered. To incorporate the homogeneous step into the simulations, the dimensionless mass transport equation is amended to:
\begin{equation}
\frac{\partial{C}_\mathrm{B}}{\partial{\tau}} = D^{'}_\mathrm{B}\left(\frac{\partial^2C_\mathrm{B}}{\partial{R^2}} + \frac{\partial^2C_\mathrm{B}}{\partial{Z^2}} + \frac{1}{R}\frac{\partial{C}_\mathrm{B}}{\partial{R}}\right) - KC_\mathrm{B}
\end{equation}
where $K$ is a dimensionless homogeneous rate constant, defined as:
\begin{equation}
K = \frac{r_\mathrm{e}^2}{D_\mathrm{A}}k
\end{equation}
and $k$ is a first order homogeneous rate constant for the reaction of species B.

\subsection{Computational Methods}

The above model is discretised using the Crank-Nicolson method\cite{Crank1947} and solved numerically over discrete spatial and temporal grids. The temporal grid is defined initially in terms of the applied dimensionless potential, $\theta$. A parameter $N_\theta$ is defined as the number of discrete time steps per unit $\theta$. The value of $\theta$ at each time step $k$ is then defined:
\begin{equation}
\theta_k = \theta_{k-1} \mp \frac{1}{N_\theta}
\end{equation}
where the minus sign is used for the reductive sweep, and the plus sign for the oxidative sweep. The dimensionless time, $\tau$, at each time step is then calculated as:
\begin{equation}
\tau_k = \tau_{k-1} + \frac{1}{N_\theta \sigma}
\end{equation}

\begin{figure}[h]
\begin{center}
\includegraphics[width = 0.8\textwidth]{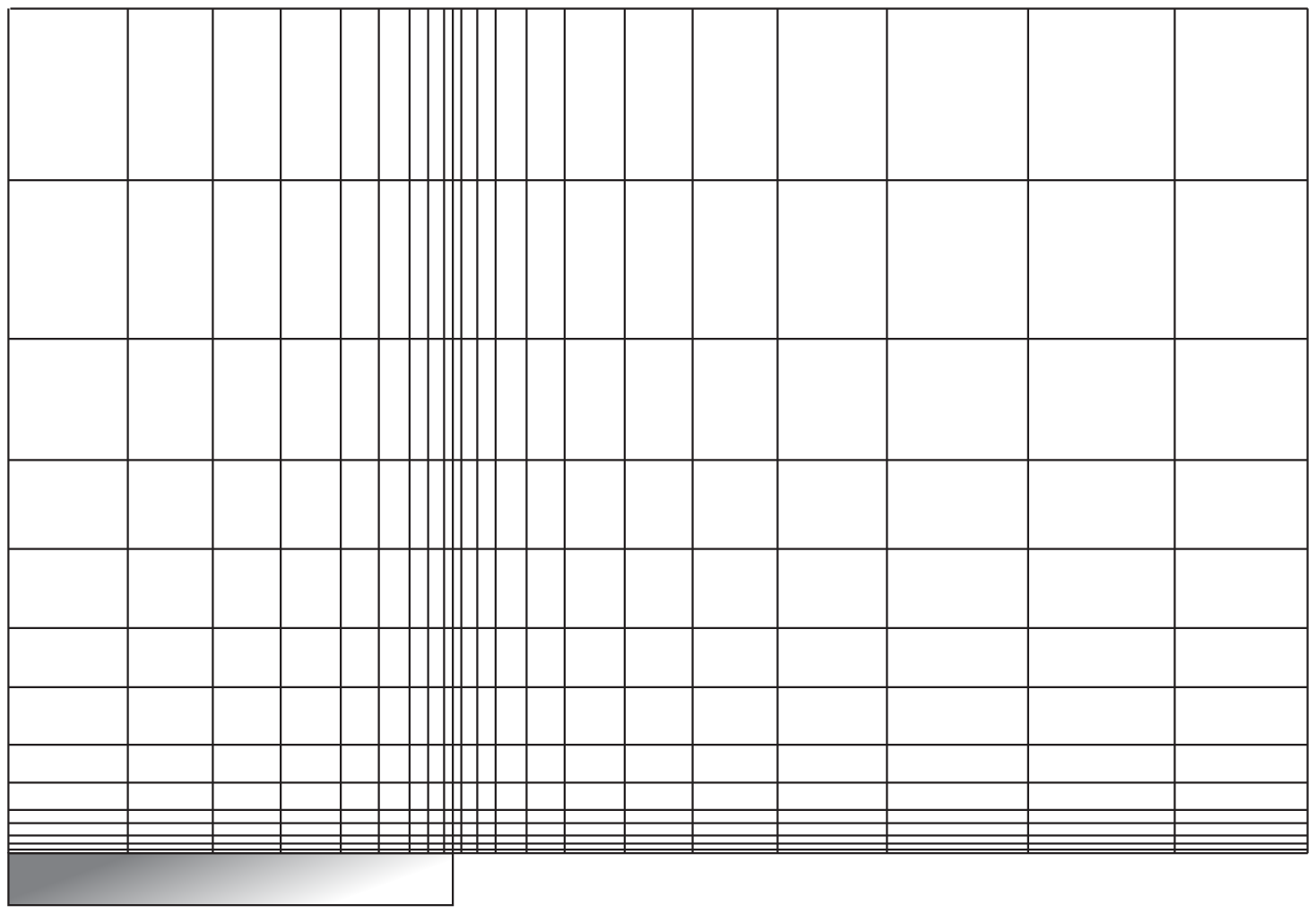}
\caption{Schematic diagram of the spatial grid used in microdisc simulations.} \label{GRID}
\end{center}
\end{figure}

A non regular, exponentially expanding spatial grid is used to ensure computational efficiency. In the Z direction, the first spatial point (at the electrode/insulating surface) is defined as $Z_0 = 0$. The first step size is defined as $\Delta$, with subsequent step sizes growing increasingly large until the simulation boundary is reached. Each $Z$ direction spatial point is defined as:
\begin{equation}
Z_j = Z_{j-1} + \gamma\left(Z_{j-1} - Z_{j-2}\right)
\end{equation}
The $R$ direction spatial grid is defined in the same way, except that rather than expanding from a simulation boundary, the grid expands away from the edge of the electrode at $R = 1$ in both directions until $R = 0$ and $R = R_\mathrm{max}$ are reached. Fig. \ref{GRID} shows a schematic diagram of some of the spatial grid points used in simulations.

In order to achieve convergence to within 0.5\% of a fully converged result, the following grid parameters were used: $N_\theta=1000$, $\Delta = 8\times 10^{-5}$ and $\gamma = 1.25$. The Alternating Direction Implicit (ADI)\cite{Britz2005} method was employed to simulate the voltammetry, in conjuction with the Thomas Algorithm\cite{Press2007} to solve the mass transport equation and boundary conditions simultaneously in matrix form. The model was coded in C++ with multithreaded parallel programming employed to increase efficiency. Typical running times were \emph{c.a.} 10 minutes on a 2.26 GHz Intel(R) Xenon(R) CPU with 2.25 GB of RAM.

\subsection{Simulating voltammetry at impacting spherical nanoparticles}

\begin{figure}[h]
\begin{center}
\includegraphics[width = 0.9\textwidth]{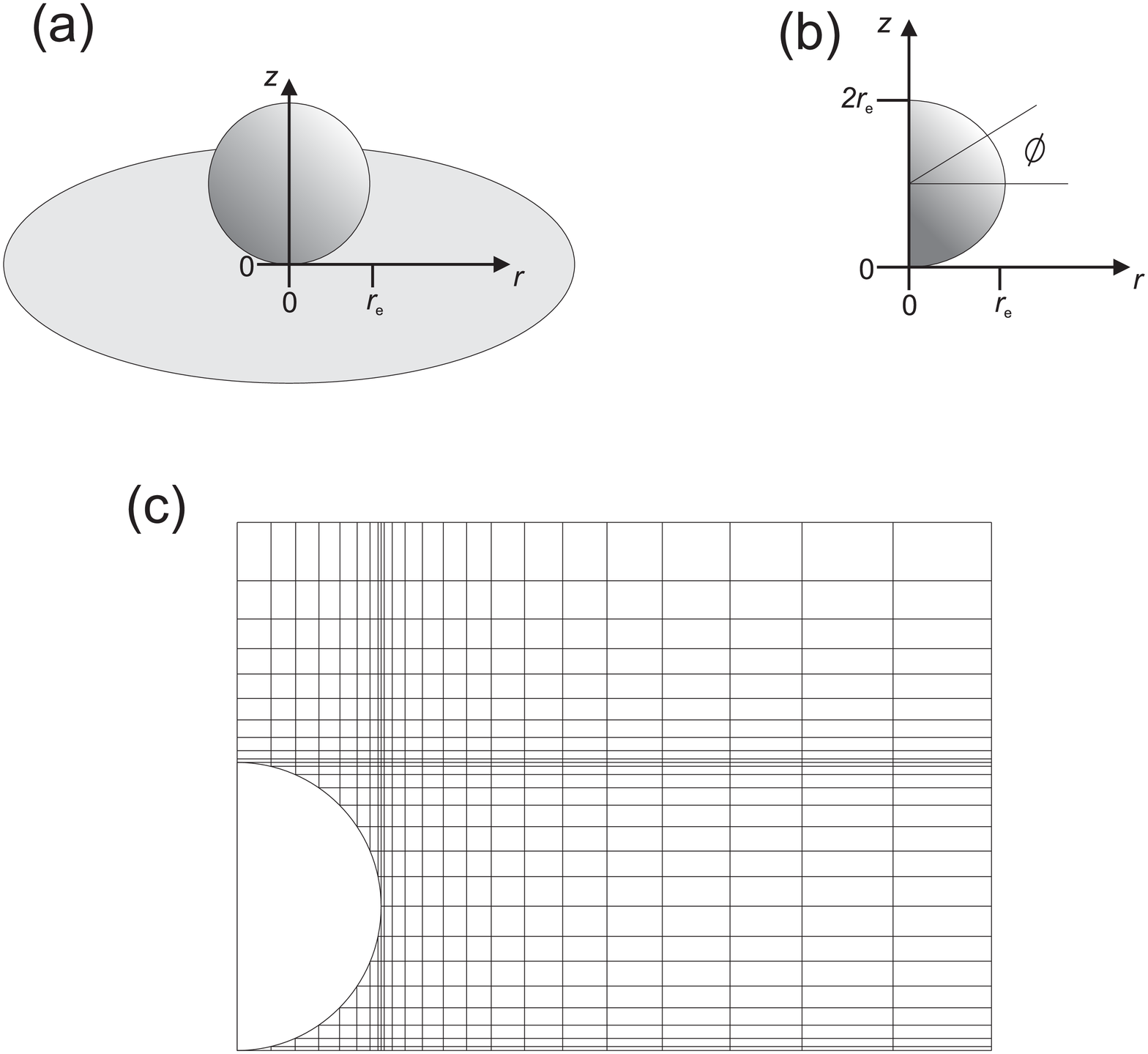}
\caption{(a) Schematic diagram of an electroactive spherical nanoparticle resting on a conductive, inert material, defining the spatial coordinate system used in simulations. (b) The two dimensional simulation space used in this study. (c) Schematic diagram of the spatial grid used in simulations.} \label{SPHERE ELECTRODE SCHEMATIC}
\end{center}
\end{figure}

Electrochemical reactions can occur at the surface of nanoparticles impacting upon a conductive, but electrochemical inert surface. A schematic diagram of a spherical nanoparticle in contact with a conducting but otherwise inert surface is shown in Fig. \ref{SPHERE ELECTRODE SCHEMATIC} (a). Upon contact, the nanoparticle becomes electrochemically active, and a species in solution can be reduced/oxidised on its surface, if a suitable potential is applied to the inter surface. This is modelled in the same way as for a microdisc described above, but with changed boundary conditions\cite{Ward2012, Molina2014, Henstridge2014, Katelhon2014}. A cylindrical coordinate system  is still used, as defined in Fig. \ref{SPHERE ELECTRODE SCHEMATIC}. Part (a) shows a schematic of the spherical electrode on the surface. Part (b) shows the 2 dimensional simulation space.

The electrode surface boundary condition is now applied at over the line $R^2 + (Z-1)^2=1$ where the flux normal to the electrode surface is given by:
\begin{equation}
\frac{\partial{C_\mathrm{A}}}{\partial{N}} = k_\mathrm{red}C_\mathrm{A}^0 - k_\mathrm{ox}C_\mathrm{B}^0
\end{equation}
where $N$ is a coordinate normal to the electrode surface, and $k_\mathrm{red}$ and $k_\mathrm{ox}$ are as defined above for either Butler-Volmer or Marcus-Hush kinetics. Other boundary conditions were zero flux conditions at the $R = 0$, $R = R_\mathrm{max} (= 1 + \sqrt{D_\mathrm{max}\tau_\mathrm{max}})$, $Z = 0$ and $Z = Z_\mathrm{max} (= 2 + \sqrt{D_\mathrm{max}\tau_\mathrm{max}})$.

The dimensionless flux is then defined as:
\begin{equation}
J = \int^{\pi/2}_{-\pi/2}\left[\frac{\partial{C_\mathrm{A}}}{\partial{R}}\text{cos}\phi + \frac{\partial{C_\mathrm{A}}}{\partial{Z}}\text{sin}\phi\right]\text{cos}\phi\text{ d}\phi
\end{equation}
The current is then defined, as for a microdisc above, as:
\begin{equation}
I = 2\pi FD_\mathrm{A}c^{*}_\mathrm{A}r_\mathrm{e}J
\end{equation}

The temporal grid for the simulation of voltammetry at a spherical electrode is the same as defined above for a microdisc electrode. To define the spatial grid, the geometry of the electrode is taken into account. The electrode surface is divided into $N_\phi$ equal spaced points, with a grid vertex at each of these points. This defines the grid points in the region $R \leq 1$ and $Z \leq 2$, as shown schematically in Fig. \ref{SPHERE ELECTRODE SCHEMATIC}. In the regions $R > 1$ and $Z > 2$, the grid expands exactly as outlined for the microdisc above. Convergence studies found $N_\phi=200$ sufficient to ensure convergence. Computational methods were identical to those employed for microdisc simulations. Simulation running times were \emph{c.a.} 10 mins  on a 2.26 GHz Intel(R) Xenon(R) CPU with 2.25 GB of RAM.

\subsection{Chronoamperometry} \label{CHRONO SECTION}

The Shoup and Szabo equation is used to calculate expected chronoamperometric responses at microdisc electrodes to within 0.6\% error\cite{Shoup1982}. For a reduction:
\begin{eqnarray}
I & = & -4nFD_\mathrm{A}c_\mathrm{A}r_\mathrm{e}f\left(\tau\right) \\
f\left(\tau\right) & = & 0.7854 + 0.4432\tau^{-0.5} + 0.2146\text{exp}\left[-0.3912\tau^{-0.5}\right] \\
\tau & = & \frac{D_\mathrm{A}}{r_\mathrm{e}^2}t
\end{eqnarray}

\subsection{Quantifying Error} \label{MSAD}

In order to examine the relative merits of Butler-Volmer theory and Marcus-Hush theory in simulating cyclic voltammetry, a quantitative measure of error is desirable. This study uses the Mean Scaled Absolute Deviation (MSAD) to quantify the error between experiment and theory. The MSAD value for a simulated fit is given by:
\begin{equation}
\text{MSAD} = \frac{1}{N}\sum_N{\left\lvert\frac{I_\mathrm{exp} - I_\mathrm{sim}}{I_\mathrm{exp}}\right\rvert} \times 100\%
\end{equation}
where $N$ is the number of experimental data points. Since it is unlikely that each experimental data point will have an exactly corresponding simulated data point, a linear interpolation is used between the two simulated data points surrounding the experimental data point in question. This interpolated simulated current is then used in the calculation of the MSAD value.

Small values of the experimental current can lead to extremely large MSAD values despite the absolute difference between experiment and theory being small. For this reason, all experimental data points with a value less than 5\% of the maximum absolute value of the current are neglected. This includes the initial part of the reductive wave as well as a narrow part of the oxidative wave, where the current passes from negative to positive values.

\section{Theoretical results}

Previous work by Feldberg\cite{Feldberg2010} and Henstridge \emph{et. al.}\cite{Henstridge2014} showed that in the Marcus-Hush model of electron transfer kinetics, steady state currents can be kinetically rather than diffusionally limited for small values of the reorganisation energy, $\lambda$. In Feldberg's study, an approximate equation of Oldham and Zoski\cite{Oldham1988} describing steady state voltammetry at a microdisc was invoked and modified to use Marcus-Hush, rather than Butler-Volmer kinetics. Henstridge used numerical simulation to generate Marcus-Hush steady state responses at a conductive nanosphere resting on an electrochemically inert but conductive surface. In this study, we simulate steady state voltammetry numerically using the symmetric Marcus-Hush formalism of electron transfer, and compare the circumstances in which a kinetically limited current is observed at a microdisc to those at a nanosphere. We also extend previous work showing the equivalence of microdiscs and spheres-on-surfaces under the Butler-Volmer formalism (subject to a mass transport correction factor)\cite{Molina2014}, and show that the approximate equivalence is retained when Marcus-Hush kinetics are used.

\subsection{Kinetically limited currents at a microdisc electrode}

\begin{figure}[h]
\begin{center}
\includegraphics[width = 1.0\textwidth]{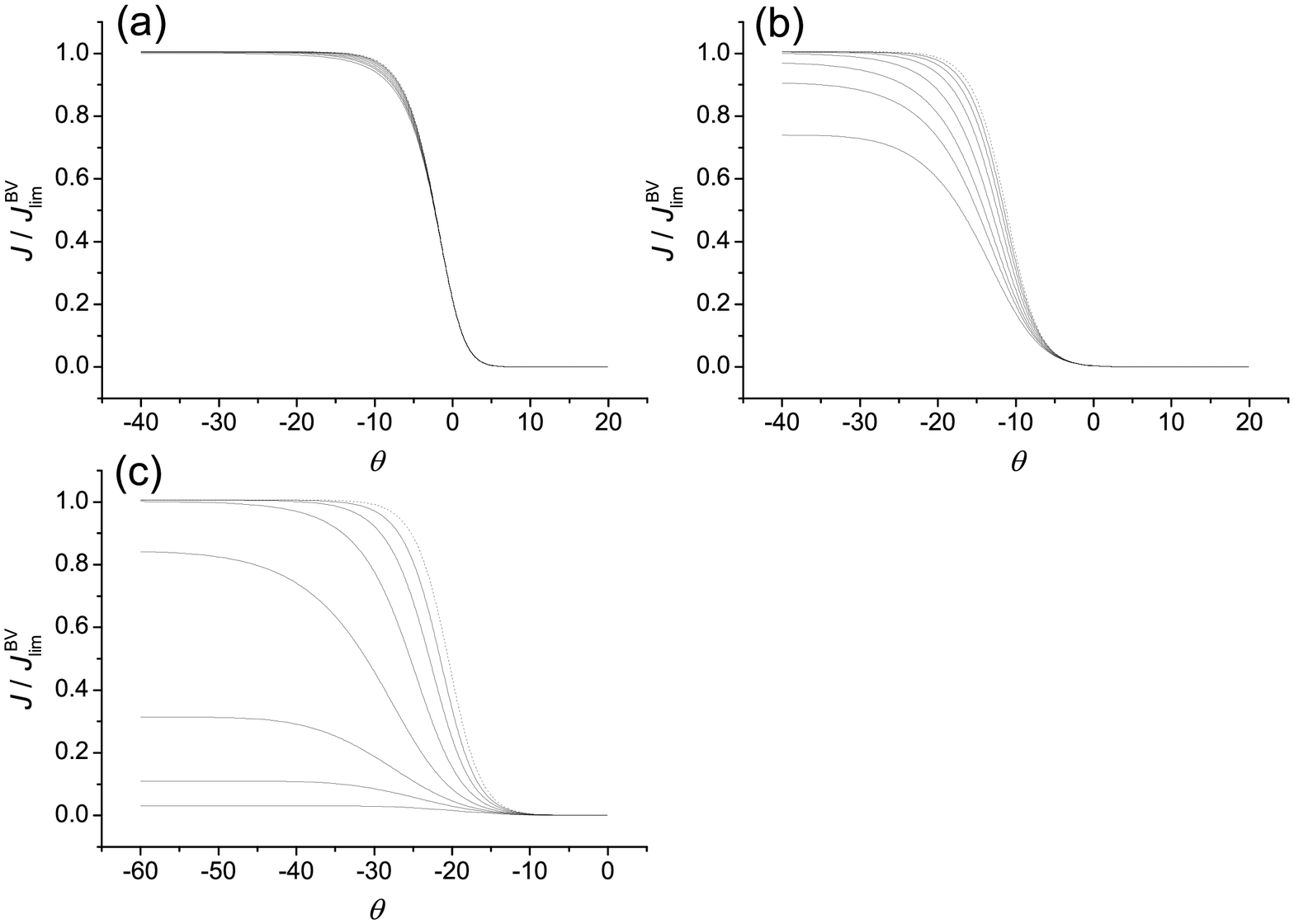}
\caption{Simulated steady state voltammetry microdisc electrode at various $\Lambda$ values and $K^0$. (a) $K^0=5\times10^{-1}$ (b) $K^0=5\times10^{-3}$ and (c) $K^0=5\times10^{-5}$. In each case the dotted line represents the Butler-Volmer result ($\alpha=0.5$). $\lambda$ values are 20 (furthest from Butler-Volmer result), 25, 30, 40, 60, 100 and 200 (closest to Butler-Volmer result).} \label{VARY K DISC}
\end{center}
\end{figure}

Steady state voltammetry at a microdisc electrode was simulated using the model outlined above. Various values of dimensionless reorganisation energy, $\Lambda$, and dimensionless heterogeneous rate constant, $K^0$, were simulated, and the results are shown in Fig. \ref{VARY K DISC}. Part (a) shows results for $K^0 = 5\times10^{-1}$, (b) shows $K^0 = 5\times10^{-3}$ and (c) shows $K^0 = 5\times10^{-5}$. Also shown in each case is the steady state voltammetry simulated using Butler-Volmer kinetics with the same values of $K^0$, and $\alpha=0.5$, and Marcus-Hush results are shown relative to this Butler-Volmer result. For all three values of $K^0$, it is seen that large $\Lambda$ values produce the same limiting current as the Butler-Volmer model. As $\Lambda$ becomes smaller, the limiting current becomes less than predicted by the Butler-Volmer model as kinetic control takes over. This effect is more extreme for the smallest values of $K^0$, where even a $\Lambda$ value of 200 has not reached the Butler-Volmer limit at all points on the wave. For $K^0=5\times10^{-1}$, unfeasibly small $\Lambda$ values would be required to show significant deviation form the Butler-Volmer result. This is in qualitative agreement with the previous studies of this nature\cite{Feldberg2010, Henstridge2014}.

The definition of $K^0$ in our dimensionless model is:
\begin{equation}
K^0 = \frac{r_\mathrm{e}k^0}{D_\mathrm{A}}
\end{equation}
Lowering $K^0$ will therefore mean that either $k^0$ or $r_\mathrm{e}$ is lowered, or $D_\mathrm{A}$ is raised. Lowering $k^0$ trivially means the electrochemical kinetics become more likely to be the limiting factor in the steady state current as opposed to diffusion. Either lower $r_\mathrm{e}$ or increasing $D_\mathrm{A}$ also both lower $K^0$ and lead to a more kinetically limited current. This is due to the fact that either of these changes will increase mass transport efficiency to the electrode surface, thus making electrochemical kinetics the more limiting factor. Therefore, in order to see a kinetically limited current, it is desirable to have a small $k^0$, a small $r_\mathrm{e}$ and a large $D_\mathrm{A}$.

\subsection{Comparison of Marcus-Hush theory at microdisc and impacting nanoparticles}

\begin{figure}[h]
\begin{center}
\includegraphics[width = 0.9\textwidth]{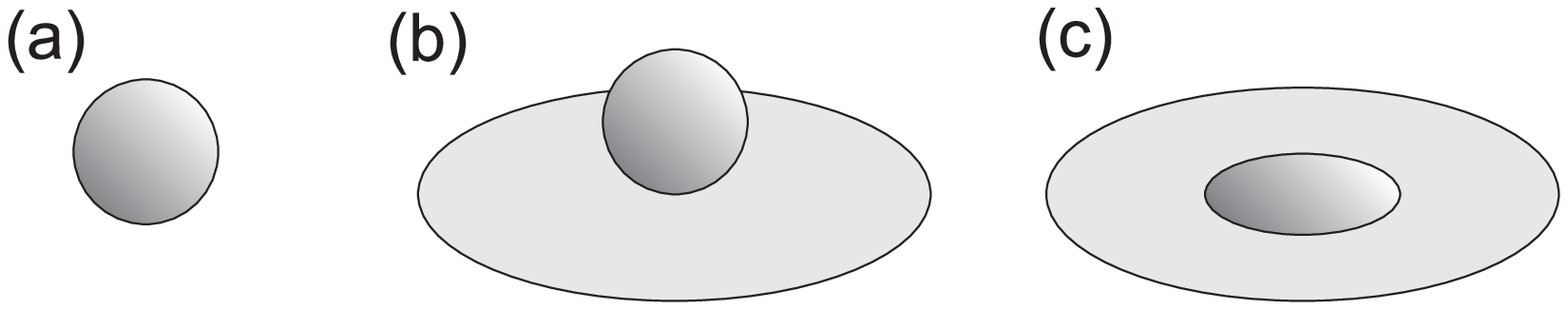}
\caption{Schematic diagram of three electrode geometries which produce steady state voltammetry described by Equation \ref{STEADY STATE EQUATION}. (a) Isolated spherical electrode, (b) spherical electrode resting on a conductive, inert surface and (c) a microdisc electrode.} \label{3 GEOMETRIES}
\end{center}
\end{figure}

Recent work by Molina \emph{et. al.}\cite{Molina2014} shows that steady state voltammetry at an isolated spherical electrodes alone in solution, a spherical electrode supported on a conductive but inert surface and a microdisc electrode (all shown schematically in Fig. \ref{3 GEOMETRIES}) may all be approximately described (under the Butler-Volmer formalism) by a single equation:
\begin{equation}
\frac{I}{I_\mathrm{lim}} = \frac{\overline{K^0}\text{exp}\left(-\alpha\theta\right)}{1 + \overline{K^0}\text{exp}\left(-\alpha\theta\right)\left[1 + \text{exp}\left(\theta\right)\right]} \label{STEADY STATE EQUATION}
\end{equation}
where $\overline{K^0}$ is a dimensionless heterogeneous rate constant scaled by a geometry dependent mass transport factor. These parameters are defined for isolated spheres (is) spheres supported on surfaces (sos) and microdiscs (md) as:
\begin{eqnarray}
\overline{K^0_\mathrm{is}} & = & \frac{k^0r_\mathrm{e}}{D_\mathrm{A}} = K^0_\mathrm{is}\\
\overline{K^0_\mathrm{sos}} & = & \frac{k^0r_\mathrm{e}}{D_\mathrm{A}}\frac{1}{\text{ln}2} = \frac{K^0_\mathrm{sos}}{\text{ln}2}\\
\overline{K^0_\mathrm{md}} & = & \frac{k^0r_\mathrm{e}}{D_\mathrm{A}}\frac{\pi}{4} = \frac{K^0_\mathrm{md}\pi}{4}
\end{eqnarray}
Therefore, steady state voltammetry at a microdisc of electrochemical rate constant $K^0_\mathrm{md}$ produces approximately the same result as steady state voltammetry at a sphere on a surface of heterogeneous rate constant $K^0_\mathrm{sos} = \pi \text{ln}\left(2\right)K^0_\mathrm{md}/4$, as was shown.

\begin{figure}[h]
\begin{center}
\includegraphics[width = 0.9\textwidth]{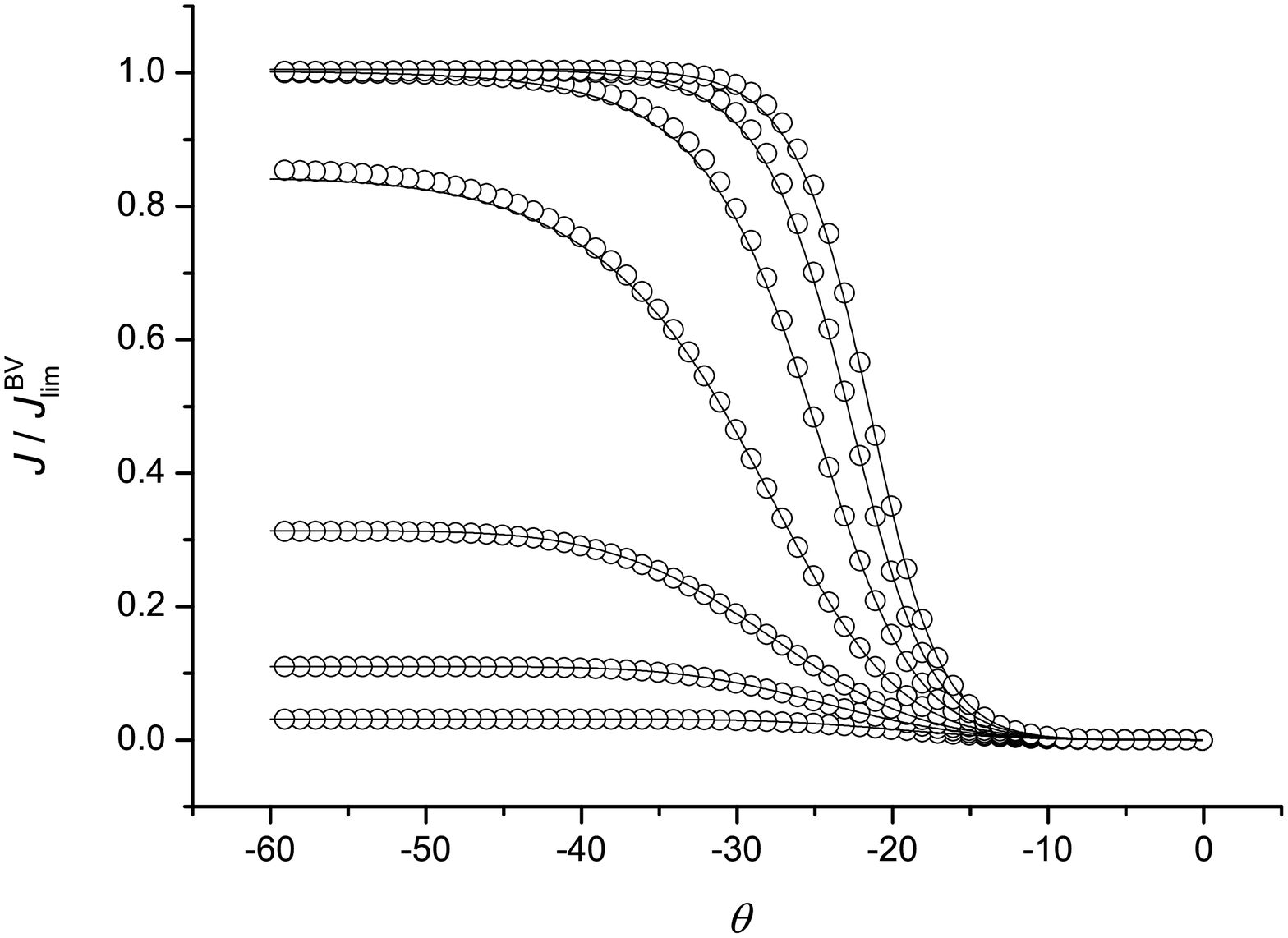}
\caption{Comparison of steady state voltammetry at a microdisc electrode (solid lines) and a sphere on a surface electrode (circles) using Marcus-Hush kinetics at various values of $\lambda$. For the microdisc electrode, $K^0 = 5 \times 10^{-5}$, and for the sphere $K^0 = \frac{\pi \text{ln}2}{4}\left(5 \times 10^{-5}\right)$. $\Lambda$ is equal to 20 (bottom result), 25, 30, 40, 60, 100 and 200} \label{COMPARE}
\end{center}
\end{figure}

We now investigate if this relation holds under the Marcus Hush formalism over a wide range of $K^0$ and $\Lambda$ values. To do so, steady state voltammetry was simulated using the Marcus-Hush model described above for both a microdisc and a sphere on a surface. For each microdisc heterogeneous rate constant ($K^0_\mathrm{md}$) that was used, a corresponding sphere on a surface simulation was run with $K^0_\mathrm{sos} = \pi \text{ln}\left(2\right)K^0_\mathrm{md}/4$, with $\Lambda$ the same in each case. Fig. \ref{COMPARE} shows simulated steady state voltammograms at a microdisc electrode (solid line) and a sphere on a surface electrode (circles) at varying values of $\Lambda$ (20, 24, 30, 40, 60, 100 and 200). $K^0$ is equal to $5\times10^{-5}$ for the microdisc and $\frac{\pi \text{ln}2}{4}\left(5 \times 10^{-5}\right)$ for the sphere. Qualitatively, the best agreement between the two geometries is seen at the smallest values of $\Lambda$ where an almost exact fit is observed. Less good fits are seen at intermediate values of $\Lambda$.

\begin{figure}[h]
\begin{center}
\includegraphics[width = 0.9\textwidth]{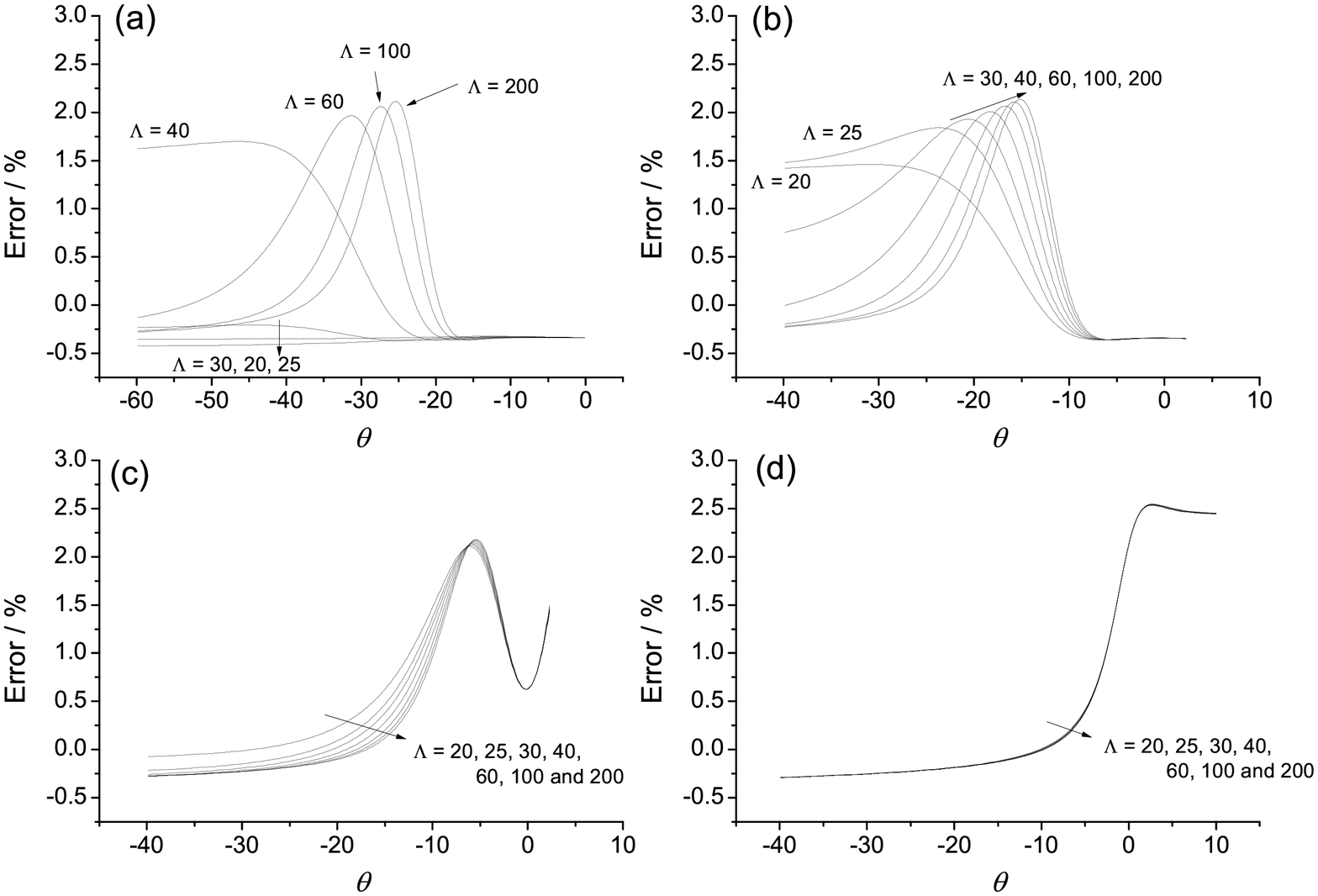}
\caption{Percentage error in the correspondence between steady state voltammetry at a microdisc and a sphere on a surface for various values of $K^0$ and $\lambda$. (a) $K^0_\mathrm{md}=5\times10^{-5}$, (b) $K^0_\mathrm{md}=5\times10^{-3}$, (c) $K^0_\mathrm{md}=5\times10^{-1}$, (d) $K^0_\mathrm{md}=5\times10^{1}$. Corresponding $K^0_\mathrm{sos}$ values are obtained by multiplying $K^0_\mathrm{md}$ by $\frac{\pi\text{ln}2}{4}$. $\Lambda$ values are indicated on the graphs.} \label{ERROR}
\end{center}
\end{figure}

Fig. \ref{ERROR} (a) shows the percentage error between the fits in Fig. \ref{COMPARE}, as well as for three other values of $K^0_\mathrm{md}$: $5\times10^{-3}$ (b), $5\times10^{-1}$ (c), and $5\times10^{1}$ (d). To obtain the corresponding values of $K^0_\mathrm{sos}$, $K^0_\mathrm{md}$ is multiplied by $\frac{\pi\text{ln}2}{4}$. The percentage error is defined as:
\begin{equation}
\text{Error} = \frac{J_\mathrm{sos}-J_\mathrm{md}}{J_\mathrm{md}} \times 100\%
\end{equation}
It is seen that as $K^0$ increases, the error becomes less and less dependent on the values of $\Lambda$. This is because at high $K^0$ the Marcus-Hush model of electron transfer approaches the Butler-Volmer limit, and becomes independent of $\Lambda$. Over the whole range of $K^0$ and $\Lambda$ values here studied, the error in the equivalence between a microdisc and a sphere on a surface never exceeds 2.55\% in regions of appreciable current. It is thus reasonable to say that the correspondence between the two geometries holds within the Marcus-Hush model of electron transfer, as well as within the Butler-Volmer model.

We now apply our Butler-Volmer and Marcus-Hush models for voltammetry at a microdisc to the reduction of oxygen in two ionic liquids, and compare the effectiveness of the models in simulating experimental data.

\section{Experimental}

\subsection{Chemical reagents}

Ferrocene (Fe(C$_5$H$_5$)$_2$, Aldrich, 98\%), tetra-n-butylammonium perchlorate (TBP, Fluka, Puriss electrochemical grade, 99\%) and acetonitrile (MeCN, Fischer Scientific, HPLC grade, 99\%) were used as received. Ionic liquids trihexyltetradecylphosphonium \emph{bis}(trifluoromethylsulfonyl)-imide ([P$_{14,6,6,6}$][NTf$_2$]) and 1-butyl-1-methylpyrrolidinium \emph{bis}(trifluoromethylsulfonyl)imide\\ ([Bmpyrr][NTf$_2$]) were kindly donated by Professor C. Hardacre of Queen's University, Belfast and were used as received. Oxygen (99.5\%) was purchased from BOC, Surrey, UK.

\subsection{Instrumentation}

Electrochemical experiments (Cyclic Voltammetry (CV) and chronoamperometry) were conducted using a $\mu$-Autolab potentiostat (Eco-Chemie, Netherlands). All experiments were conducted inside a temperature controlled Faraday cage.\cite{Evans2004} The working microdisc electrode (either Gold or Platinum, IJ Cambria Scientific Ltd, UK), 10 $\mu$m nominal diameter, was polished prior to use using a water-alumina slurry (1, 0.3, 0.05 $\mu$m, five minutes on each grade) on soft lapping pads (Buehler, Illinois).\cite{Cardwell1996} Precise radii were determined through calibration of the electrode with a 2.0 mM solution of ferrocene in acetonitrile containing 0.1 M TBAP (silver wire as both a counter and quasi-reference electrode); chronoamperometry was recorded at 298 K, and assuming a diffusion coefficient of $2.3 \times 10^{-9}$ m$^2$ s$^{-1}$,\cite{Rodgers2008b} the data was analysed with respect to the Shoup and Szabo equation\cite{Shoup1982} (described in Section \ref{CHRONO SECTION}). This gave electrode radii of $5.10 \pm 0.05$ $\mu$m (Gold) and $5.40 \pm 0.05$ $\mu$m (Platinum). A 0.5 mm silver wire was used both as a counter and a quasi-reference electrode. The experiments conducted with a gold electrode were arranged in a glass T-cell, as described previously,\cite{Rodgers2008c} with a plastic collar attached to the working electrode. A 10 $\mu$L aliquot of ionic liquid was transferred into the T-cell, which was left under vacuum (0.2 mbar) overnight, to ensure the complete removal of volatile substances, including atmospheric gases and any aqueous impurities.  In the case of the platinum working electrode, a three electrode (with two silver wires acting as a counter and a reference electrode) system was used in a sealed sample vial with a pre-vacuumed ionic liquid. Oxygen was then introduced to the T-cell or sample vial via a drying column of molecular sieves (activated at 200 $^\circ$C overnight, 4 \AA, Sigma Aldrich). The gas was left on at a flow rate of 200 - 300 cm$^3$ min$^{-1}$ for at least half an hour to reach saturation, at which point cyclic voltammograms were recorded to ensure equilibrium had been reached and the reductive current in successive experiments was stable. A wait time of minimum 15 minutes was imposed between each experiment to ensure a stable concentration of oxygen at the electrode surface.

\section{Results and discussion} \label{RESULTS AND DISCUSSION}

This section reports the comparative outcome of the use of BV theory and SMH theory to model experimental cyclic voltammetry of the reduction of oxygen in the ionic liquids [Bmpyrr][NTf$_2$] and [P$_{14,6,6,6}$][NTf$_2$]:
\begin{equation}
\mathrm{O}_2 + \mathrm{e^-} \rightleftharpoons \mathrm{O}_2^{\bullet -}
\end{equation}
This is followed by the inference of physical implications from the application of SMH and the determination of a value for solvent reorganisation energy, $\lambda$, for both RTILs.

\subsection{Reduction of Oxygen in [Bmpyrr][NTf$_2$] on a $\mu$Au electrode} \label{O2 IN BMPYRR}

\begin{figure}[h]
\begin{center}
\includegraphics[width = 0.9\textwidth]{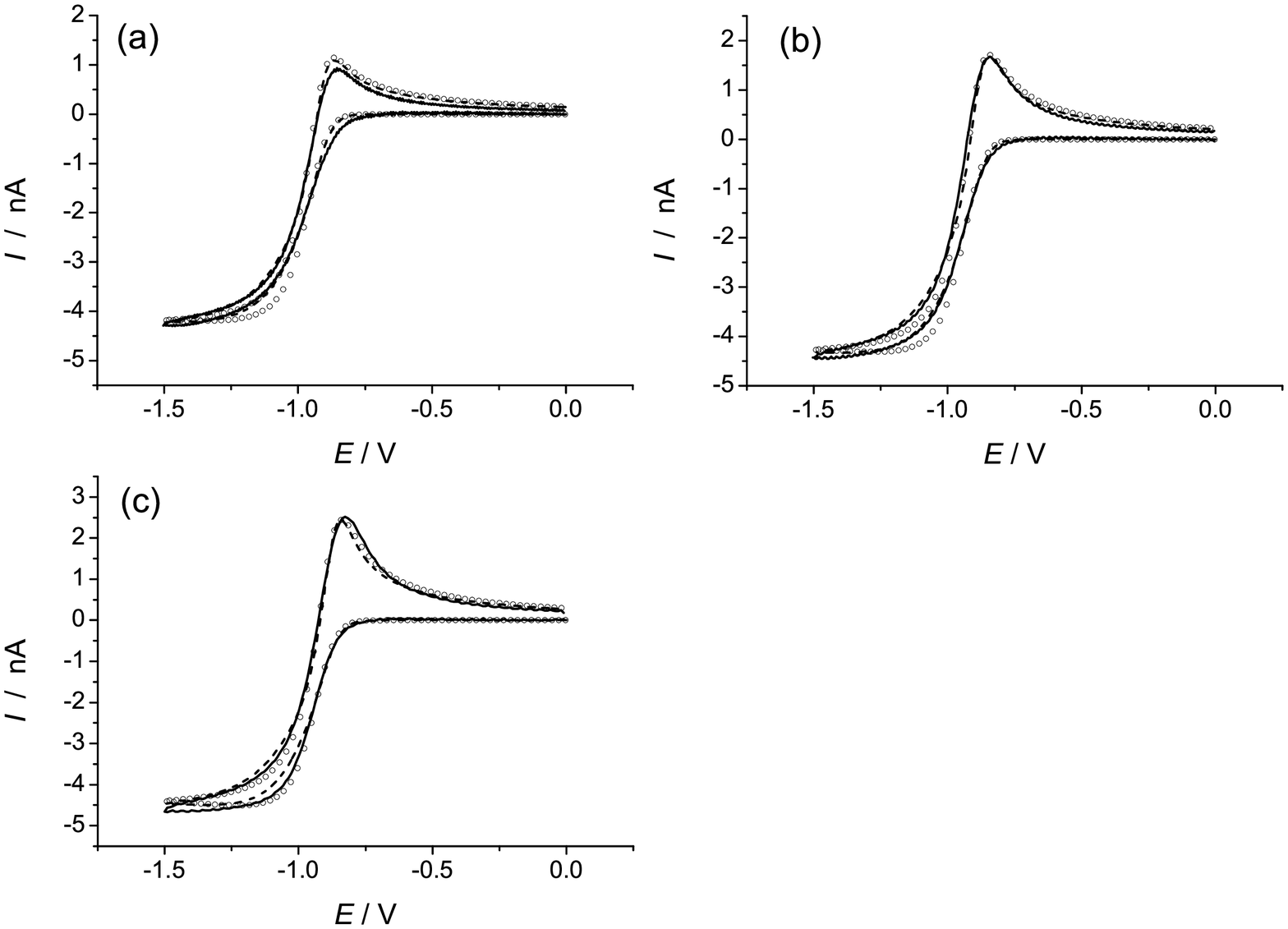}
\caption{Experimental cyclic voltammetry of the reduction of oxygen in [Bmpyrr][NTf$_2$] on a $\mu$Au electrode (solid line) at 298 K, compared to Butler-Volmer Theory (dashed line), and symmetric Marcus-Hush Theory (circles) for scan rates of (a) 100 mV s${-1}$, (b) 200 mV s$^{-1}$, and (c) 400 mV s$^{-1}$. Parameters corresponding to each simulation are detailed in Table \ref{BMPYRR PARAMETERS}.} \label{BMPYRRNTF2}
\end{center}
\end{figure}

The first system under study is the reduction of oxygen in [Bmpyrr][NTf$_2$]. This ionic liquid was chosen for its relatively low viscosity (70 cP at 298 K)\cite{Okoturo2004} and previously observed well characterised voltammetry for oxygen reduction.\cite{Huang2009} Cyclic voltammetry (CV) was performed on saturated solutions of oxygen in [Bmpyrr][NTf$_2$] over a range of scan rates (100-400 mV s$^{-1}$). A blank CV, from -3 V - 2 V vs Ag, was recorded prior to the introduction of oxygen to ensure the ionic liquid was free from impurities, and to estimate solution capacitance for each scan rate. The capacitative current was subtracted from the experimental current to generate corrected voltammograms prior to further analysis. In all further experimental cases, the potential was swept from 0.00 V (vs Ag), a voltage at which \emph{ca.} zero Faradaic current flows, to a potential cathodic of the reduction of oxygen (-1.50 V vs Ag), then back to 0.00 V (vs Ag). At a scan rate of 200 mV s$^{-1}$, the oxidative peak current at -0.8 V was 1.7 nA, the limiting current was -4.4 nA, and the half peak potential was -0.9 V vs Ag wire. The peak corresponding to the reduction of oxygen appeared on the forward scan at \emph{ca.} -1.3 V vs Ag, whilst the oxidative peak appeared on the reverse scan at \emph{ca.} -0.8 V vs Ag. The resultant voltammograms (Fig. \ref{BMPYRRNTF2}, a-c) show more steady-state-like behaviour in the forward wave, which corresponds to the reduction of oxygen to superoxide, than in the back wave, corresponding to the reoxidation of superoxide to oxygen. The disparity in diffusion coefficients of the reactant and product in the ionic solvent results in different types of diffusion being observed at the electrode surface. In the case of oxygen, we see convergent Fickian diffusion of oxygen, whereas for the more slowly diffusing superoxide we observe planar Fickian diffusion.\cite{Buzzeo2003}

CVs were modelled using a previously reported program\cite{Buzzeo2003} that utilises Butler-Volmer kinetics, as outlined above. As this is a well-reported system,\cite{Evans2004, Katayama2005, Yaun2014,Monaco2012} the starting point for the diffusion coefficient and the concentration of oxygen were taken from literature. Optimised values for $k^0$, $\alpha$, $c_\mathrm{O_2}$, $D_\mathrm{O_2}$, $D_\mathrm{O_2^{\bullet -}}$, and $E_\mathrm{f}^\minuso$ (tabulated in Table \ref{BMPYRR PARAMETERS}) were obtained to give the ``best fit'', which was quantified with the calculation of a Mean Scaled Absolute Deviation (MSAD, as described in Section \ref{MSAD}). These values were used across the scan rates with all values remaining constant, with the exception of $E_\mathrm{f}^\minuso$, which was allowed to deviate slightly ($\pm 0.02$ V) to account for the use of a quasi-reference electrode.\cite{Gritzner1984} The optimised values and the mean MSADs across the scan rates are reported in Table \ref{BMPYRR PARAMETERS}.

The value of $D_\mathrm{O_2^{\bullet -}}$, the diffusion coefficient of the superoxide ion, is an order of magnitude smaller than that of the diffusion coefficient for oxygen, $D_\mathrm{O_2}$. This is due to the charged superoxide ion's increased interaction with the ionic solvent, which reduces its movement through the solution. This behaviour has been reported in other ionic liquids by Buzzeo \emph{et al},\cite{Buzzeo2003} Evans  \emph{et al},\cite{Evans2004} and Huang  \emph{et al}.\cite{Huang2009} The value for the transfer coefficient, $\alpha$, of 0.29, suggests an early, reactant-like transition state.\cite{Compton2010} Another notable outcome is the value for $k^0$, of $1.5 \times 10^{-3}$ cm s$^{-1}$, which is comparable to $k^0$ for oxygen reduction  in other ionic solvents, as seen in Table \ref{LIT VALUES}, which details transfer coefficients and rate constants for oxygen reduction in a range of ionic liquids.

Symmetric Marcus-Hush (SMH) theory was next used to model the electrode kinetics, and the same procedure as outlined above was followed to establish independent, optimal values for the parameters of concern. In this case, the transfer coefficient, $\alpha$, is no longer a parameter that is taken into consideration. Instead, the reorganisation energy, $\lambda$, is used. The standard electrochemical rate constant, $k^0$, also takes on different values in this model. The best fit values and the mean MSAD values are outlined in Table \ref{BMPYRR PARAMETERS}. 

The experimental value for $\lambda$ was determined to be 0.4 $\pm$ 0.1 eV in this solvent. This was determined by establishing best-fit parameters for the SMH model with experimental data. The BV program reproduces experimental data with a greater level of accuracy that the SMH program, as evidenced by lower MSAD values and a better fit, as seen in Table \ref{BMPYRR PARAMETERS} and Fig. \ref{BMPYRRNTF2}. This suggests BV kinetics parameterise the data better than the SMH model. A similar conclusion has been reached with respect to the modelling of redox processes in molecular solvents\cite{Suwatchara2011, Suwatchara2012}.  

\subsection{Reduction of Oxygen in [P$_{14,6,6,6}$][NTf$_2$] on a $\mu$Pt electrode.} \label{P14666NTf2 RESULTS}

The second system under consideration is the chemically irreversible reduction of oxygen in [P$_{14,6,6,6}$][NTf$_2$]. This ionic liquid has high hydrophobicity\cite{O'Mahony2008} and previously reported well defined voltammetry.\cite{Evans2004} Cyclic voltammetry (CV) was performed on saturated solutions of oxygen in [P$_{14,6,6,6}$][NTf$_2$] over a range of scan rates (100-400 mV s$^{-1}$) (see Fig. \ref{P14666NTF2}, a-c). A blank CV was recorded prior to the introduction of oxygen to ensure the ionic liquid was free from impurities. The baselines of the raw experimental CVs were adjusted to generate corrected voltammograms prior to further analysis. In all cases, the potential was swept from 0.00 V vs Ag, to a potential cathodic of the reduction of oxygen (-2.3 V vs Ag). At a scan rate of 200 mV s$^{-1}$, the limiting current was -6.5 nA, and the half peak potential was -1.6 V vs Ag. The reduction of oxygen occurred on the forward scan at -2.20 V vs Ag, whilst an oxidative feature appeared on the reverse scan at -0.86 V vs Ag.

Unlike the case discussed in Section \ref{O2 IN BMPYRR}, in this situation the highly reactive superoxide that is generated by the initial reduction encounters a proton source in the form of the cation of the solvent. Evans \emph{et. al}\cite{Evans2004} detailed the mechanism of this process (shown in Scheme 1) and suggested that the observed back peak in the experimental voltammetry could be an oxygen-derived species created in follow-up reactions, most likely HO$_2^{\bullet}$. Given the documented chemical  irreversibility of this system, only the forward wave was considered for further analysis. Both the Butler-Volmer and Marcus-Hush programs were modified to account for a homogeneous chemical step in the mass transport equations, as detailed in the Theory section.

\begin{figure}[t]
\begin{center}
\includegraphics[width = 0.75\textwidth]{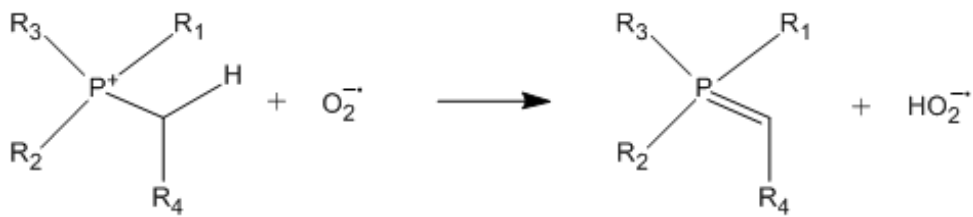}
\caption*{Scheme 1} \label{DEPROTONATION}
\end{center}
\end{figure}

\begin{figure}[h]
\begin{center}
\includegraphics[width = 0.9\textwidth]{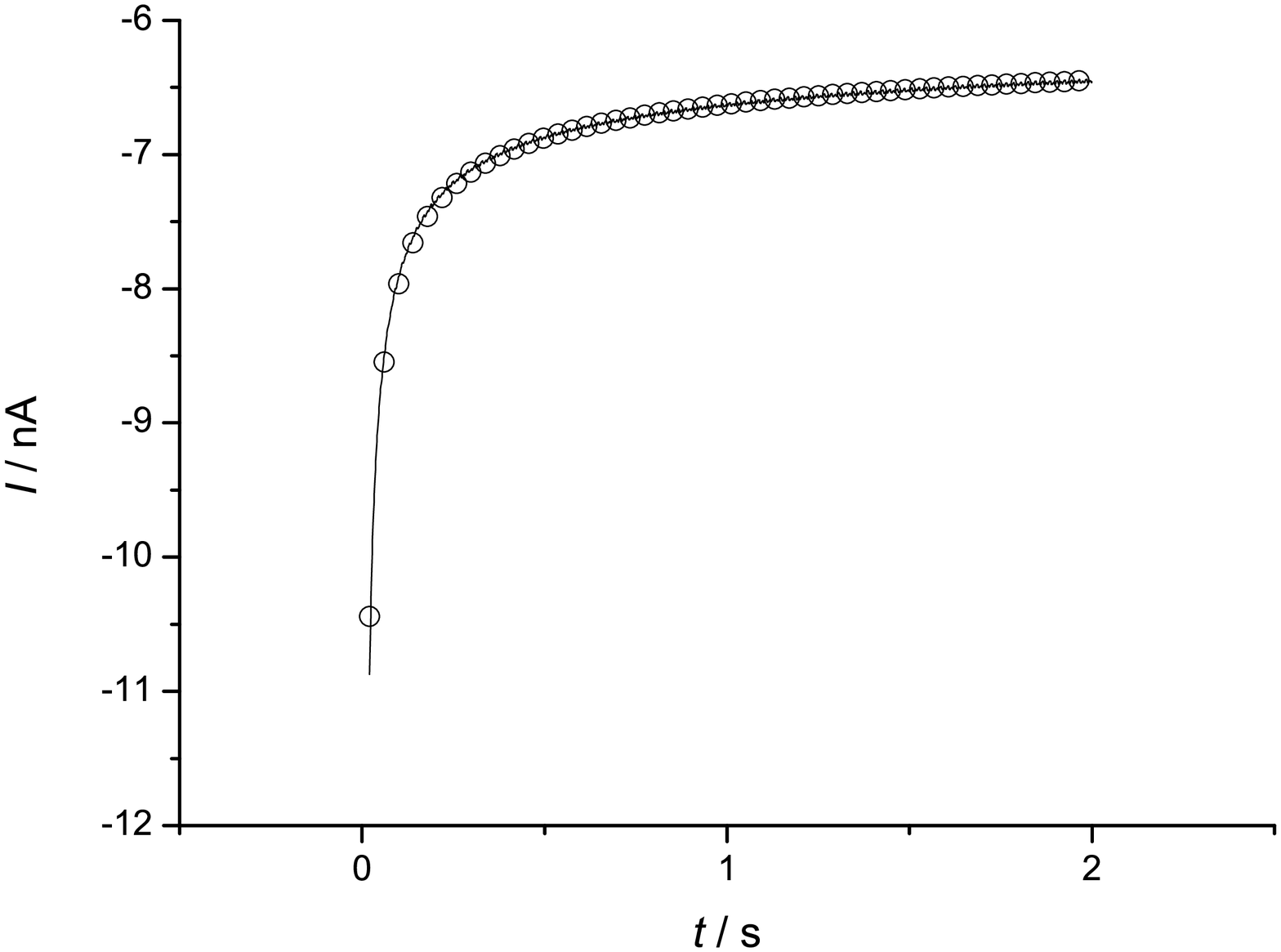}
\caption{Representative experimental chronoamperometry of the reduction of oxygen in [P$_{14,6,6,6}$][NTf$_2$] on a $\mu$Pt electrode (solid line) compared with fitting obtained using the Shoup-Szabo analysis (circles).} \label{CHRONO}
\end{center}
\end{figure}

For the purposes of determining a starting point for $D_\mathrm{O_2}$ and concentration, potential step chronoamperometry was performed on this system by stepping from a potential at which no current flows, 0.0 V, to -2.0 V for 2s (see Fig. \ref{CHRONO}). This data was then analysed with respect to the Shoup and Szabo\cite{Shoup1982} equation (as detailed in the Section 2.4), to simultaneously determine the concentration and diffusion coefficient of oxygen (see Table \ref{P14666 PARAMETERS}).

Optimised values for $k^0$, $\alpha$, $c_\mathrm{O_2}$, $D_\mathrm{O_2}$, and $E_\mathrm{f}^\minuso$ were obtained to give the ``best fit'' using the modified Butler-Volmer program described above, which was quantified with the calculation of a Mean Scaled Absolute Deviation (MSAD, as described in the Section \ref{MSAD}). These values were used across the scan rates with all values remaining constant, with the exception of $E_\mathrm{f}^\minuso$, which was allowed to deviate slightly ($\pm$ 0.02 V) to account for the use of a quasi-reference electrode. The optimised values and the mean MSADs across the scan rates are reported in Table \ref{P14666 PARAMETERS}. The values obtained for $D_\mathrm{O_2}$ and $c_\mathrm{O_2}$ are consistent with previous literature reports and are in agreement with reported diffusion coefficients and saturated concentrations of oxygen in a range of ionic liquids (Table \ref{LIT VALUES}). 

The value for the transfer coefficient, $\alpha$, of 0.31, suggests an early, reactant-like transition state. This is congruent with the result found in section \ref{O2 IN BMPYRR}, and a similar $k^0$ of $1 \times 10^{-3}$ cm s$^{-1}$ is reported, which is again in agreement with the literature values reported in Table \ref{LIT VALUES}.

SMH theory was used independently to generate the same set of parameters as generated in the BV theory with the transfer coefficient, $\alpha$, once again being replaced by a representation for reorganisation energy, $\lambda$. The experimental value for $\lambda$ was determined to be 0.5 $\pm$ 0.1 eV in this solvent. 

Once again, the BV program provides parameters with a significantly greater level of certainty that the SMH program. And, although the SMH theory gives greater physical insight compared with the phenomenological BV theory, this greater uncertainty leaves BV as the preferred theory, in terms of attaining more reliable parameterisation. 

\subsection{Comparison of Oxygen Reduction in Pyrrolidinium and Phosphonium based Ionic Liquids}

Both systems under study, oxygen reduction in either a pyrrolidinium or phosphonium based IL, display similarly low diffusion coefficients ($D_\mathrm{O_2}$ = 2.05 and 3.95 $\times 10^{-10}$ m$^2$ s$^{-1}$ respectively) and solubilities of oxygen ($c_\mathrm{O_2}$ = 10 and 7.3 mM). In addition to this, both the transfer coefficients ($\alpha$ = 0.29 $\pm$ 0.01, and 0.31 $\pm$ 0.01 respectively) and the reorganisation energies ($\lambda$ = 0.4 $\pm$ 0.1, and 0.5 $\pm$ 0.1 respectively) are similar. As outlined in the introduction, $\lambda$ encompasses contributions from both inner-sphere reorganisation of the molecule itself, and outer-sphere solvent reorganisation about the electroactive species. The inner-sphere reorganisation is due to the change in bond length on going from O$_2$ to O$_2^{\bullet -}$ (0.12 \AA)\cite{Hartnig2002}, with outer-sphere solvent reorganisation dominating in aqueous systems. Literature reports suggest the overall reorganisation energy for the oxygen self-exchange reaction in water is 1.26-4.51 eV.\cite{McDowell1984} The inner-sphere contribution was estimated to be  0.4-0.6 eV.\cite{Hartnig2002} Given the close agreement with previously reported values for inner sphere reorganisation\cite{Eberson1987}, the experimental values for $\lambda$ in both of the systems considered in this work are consistent with a process dominated by an inner sphere contribution. Previous reports have indicated that oxygen itself had weak interactions with an ionic liquid,\cite{Anthony2002} and the result reported in the present work suggests that the ionic liquid does not undergo any significant reorganisation on moving from oxygen to superoxide. This may be related to the very small size of the O$_2$/O$_2^{\bullet -}$ species compared to the component ions of the RTIL.

\begin{figure}[h]
\begin{center}
\includegraphics[width = 0.9\textwidth]{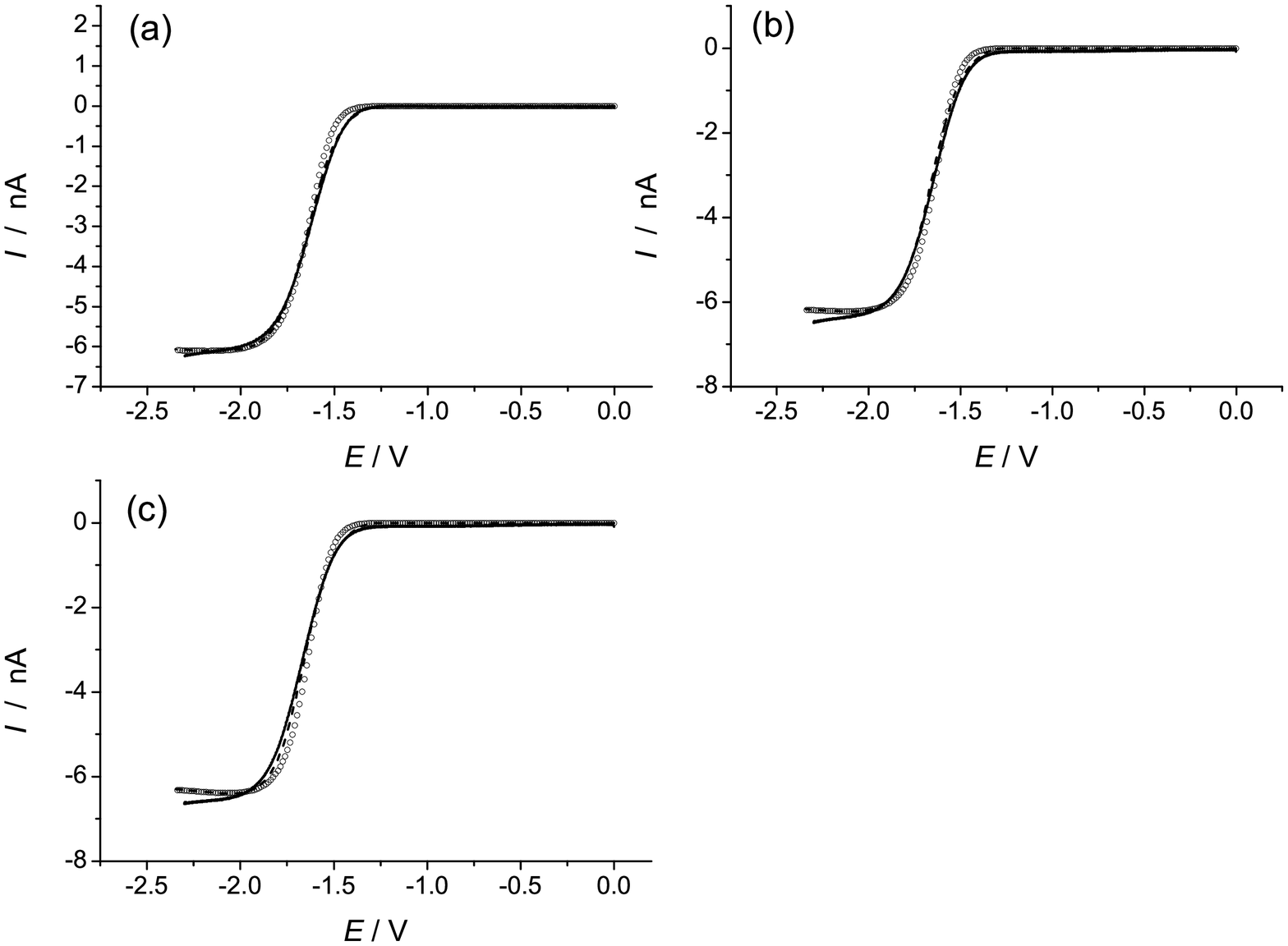}
\caption{Experimental cyclic voltammetry of the reduction of oxygen in [P$_{14,6,6,6}$][NTf$_2$] on a $\mu$Pt electrode (solid line) at 298 K, compared to Butler-Volmer Theory (dashed line), and symmetric Marcus-Hush Theory (circles) for scan rates of (a) 100 mV s$^{-1}$, (b) 200 mV s$^{-1}$, and (c) 400 mV s$^{-1}$. Parameters corresponding to each simulation are detailed in Table \ref{P14666 PARAMETERS}.} \label{P14666NTF2}
\end{center}
\end{figure}

The reported best fit values for the rate constant given by SMH theory are similar to those established by BV theory, however there is less agreement with the experimental data than is seen when the BV program is used. This is reflected both in the larger mean MSAD values, and qualitatively observing the fit as seen in Figs. \ref{BMPYRRNTF2} and \ref{P14666NTF2}. Further studies may involve consideration of asymmetric MH theory, which relaxes the requirement that the potential energy parabola are of equal curvature, and introduces a new parameter to describe the disparity in curvature between the reactant and product. This has been considered in a number of molecular systems,\cite{Laborda2012, Henstridge2012e, Laborda2013} and has been determined to fit experimental data more accurately than SMH.

\section{Conclusions}

Models were developed to simulate cyclic voltammetry at a microdisc electrode and a sphere on a surface electrode using both Butler-Volmer and symmetric Marcus-Hush kinetics. The microdisc model was used to investigate the possibility of kinetically limited steady state currents at a microdisc electrode, and it was found that this phenomenon is more likely to be observed at nano-sized sphere on a surface electrodes (\emph{e.g.} an impacting nanoparticle) because of the smaller size. It was also found that the steady state equivalence (subject to a mass transport correction) between microdisc electrodes and spherical electrodes resting on a surface also hold when Marcus-Hush kinetics are employed.

The reduction of oxygen at a microdisc electrode in two room temperature ionic liquids, [Bmpyrr][NTf$_2$] and [P$_{14,6,6,6}$][NTf$_2$], has been studied with cyclic voltammetry at 298 K. BV theory and SMH theory were used to create simulation programs, and the simulated voltammograms were compared with experimental data. BV theory, although empirical, provided voltammograms with lower MSAD values than those generated using SMH theory. A reorganisation energy, from the SMH program, was determined to be 0.4-0.5 eV for oxygen reduction in the two RTILs studied. This low value likely corresponds to a dominant inner sphere reorganisation, and limited solvent reorganisation.

Further studies to investigate whether AMH theory is better than SMH in its ability to parameterise experimental data in ionic solvents will likely be valuable.

\section*{Acknowledgments}

For funding, EELT thanks the Clarendon Fund and St John's College, Oxford, LX thanks Honeywell Analytics, and EOB thanks EPSRC and St John's College, Oxford.

\clearpage

\section*{Figures}

\clearpage

\clearpage

\clearpage

\clearpage

\clearpage

\clearpage

\clearpage

\clearpage

\clearpage

\clearpage

\clearpage

\clearpage

\clearpage

\section*{Tables}

\clearpage

\begin{table}
\begin{center}
\begin{tabular}{l l l}
\hline
Parameter & Description & Units\\
\hline
$\alpha$ & Transfer coefficient & Unitless\\
\\
$c_\mathrm{i}$ & Concentration of species i & mol m$^{-3}$\\
\\
$c_\mathrm{i}^{*}$ & Bulk solution concentration of species i & mol m$^{-3}$\\
\\
$c_\mathrm{i}^{0}$ & Electrode surface concentration of species i & mol m$^{-3}$\\
\\
$D_\mathrm{i}$ & Diffusion coefficient of species i & m$^2$ s$^{-1}$ \\
\\
$E$ & Applied potential & V\\
\\
$E_f^\minuso$ & Formal potential of A/B couple & V\\
\\
$F$ & Faraday constant = 96485 & C mol$^{-1}$\\
\\
$I$ & Current & A\\
\\
$k$ & (Pseudo) first order homogeneous rate constant & s$^{-1}$\\
\\
$k^0$ & Electrochemical rate constant & m s$^{-1}$\\ 
\\
$k_\text{red/ox}$ & Reduction/oxidation rate constant & Unitless\\
\\
$\lambda$ & Reorganisation energy & eV\\
\\
$\nu$ & Scan rate & V s$^{-1}$\\
\\
$R$ & Gas constant = 8.314 & J K$^{-1}$ mol$^{-1}$\\ 
\\
$r$ & Radial coordinate & m\\
\\
$r_e$ & Radius of electrode & m\\
\\
$T$ & Temperature & K\\
\\
$t$ & Time & s\\
\\
$z$ & $z$ coordinate & m\\ 
\hline
\end{tabular}
\end{center}
\caption{List of symbols}
\label{DIMENSIONAL}
\end{table}

\clearpage

\begin{table}
\begin{center}
\begin{tabular}{c c}
Dimensionless Parameter & Definition \\
\hline
\\
$C_\mathrm{i}$ & $\frac{c_\mathrm{i}}{c_\mathrm{A}^*}$ \\
\\
$D_\mathrm{i}^{'}$ & $\frac{D_\mathrm{i}}{D_\mathrm{A}}$ \\
\\
$J$ & $\frac{I}{2\pi FD_\mathrm{A}c_\mathrm{A}^{*}r_\mathrm{e}}$\\
\\
$K$ & $\frac{r_\mathrm{e}^2}{D_\mathrm{A}}k$\\
\\
$K^0$ & $\frac{r_e}{D_\mathrm{A}}k^0$\\
\\
$\Lambda$ & $\frac{F}{RT}\lambda$\\
\\
$R$ & $\frac{r}{r_e}$\\
\\
$\theta$ & $\frac{F}{RT}\left(E - E_f^\minuso\right)$\\
\\
$\tau$ & $\frac{D_\mathrm{A}}{r^\mathrm{2}_\mathrm{e}}t$ \\
\\
$Z$ & $\frac{z}{r_e}$\\
\\
\hline
\end{tabular}
\end{center}
\caption{Dimensionless parameters. Species A refers to the species initially present in solution before the experiment/simulation begins.}
\label{DIMENSIONLESS}
\end{table}

\clearpage

\begin{table}
\begin{center}
\begin{tabular}{c c c}
Parameter & Butler-Volmer & Marcus-Hush \\
\hline
\\
$D_\mathrm{O_2}$ / m$^2$ s$^{-1}$ & $2.05 \times 10^{-10}$ &  $2.05 \times 10^{-10}$\\
\\
$D_\mathrm{O_2^{\bullet -}}$ / m$^2$ s$^{-1}$ & $1.80 \times 10^{-11}$ &  $1.55 \times 10^{-11}$\\
\\
$c_\mathrm{O_2}$ / mM & 10 & 10 \\
\\
$k^0$ / cm s$^{-1}$ & 0.0023 & 0.0015 \\
\\
$E_\mathrm{f}^\minuso$ / V & $-0.87 \pm 0.02$ & $-0.87 \pm 0.02$ \\
\\
$\alpha$ & 0.29 & - \\
\\
$\lambda$ / eV & - & 0.4\\
\\
MSAD (average) / \% & 9.52 & 13.9\\
\hline
\end{tabular}
\end{center}
\caption{Simulation Results for the reduction of oxygen in [Bmpyrr][NTf$_2$] on a $\mu$Au electrode.}
\label{BMPYRR PARAMETERS}
\end{table}

\clearpage

\begin{table}
\begin{center}
\begin{tabular}{c c c c c c}
Ionic Liquid & $c_\mathrm{O_2}$ / mM & $D_\mathrm{O_2}$ / m$^2$ s$^{-1}$ & $D_\mathrm{O_2^{\bullet -}}$ / m$^2$ & $k^0$ / cm s$^{-1}$& $\alpha$\\
\hline
\\
{[}Emim][BF$_4$]\cite{Zhang2004} & $1.1 \pm 0.2$ & $1.7 \pm 0.2 \times 10^{-9}$ & - & $0.94 \pm 0.13 \times 10^{-3}$ & $0.46 \pm 0.02$\\
\\
{[}Pmim][BF$_4$]\cite{Zhang2004} & $0.97 \pm 0.05$ & $1.3 \pm 0.2 \times 10^{-9}$ & - & $1.5 \pm 0.4 \times 10^{-3}$ & $0.36 \pm 0.03$\\
\\
{[}Bmim][BF$_4$]\cite{Zhang2004} & $1.1 \pm 0.1$ & $1.2 \pm 0.1 \times 10^{-9}$ & - & $0.83 \pm 0.2 \times 10^{-3}$ & $0.36 \pm 0.02$\\
\\
{[}MOPMPip][NTf$_2$]\cite{Hayyan2011} & $14.3$ & $1.0 \times 10^{-10}$ & - & - & $0.18$\\
\\
{[}HMPyrr][NTf$_2$]\cite{Hayyan2011} & $14.5$ & $2.5 \times 10^{-10}$ & - & - & $0.13$\\
\\
{[}Emim][NTf$_2$]\cite{Buzzeo2003} & $3.9$ & $8.3 \times 10^{-10}$ & $1.27 \times 10^{-10}$ & - & -\\
\\
{[}N$_{6,2,2,2}$][NTf$_2$]\cite{Buzzeo2003} & $11.6$ & $1.48 \times 10^{-10}$ & $4.66 \times 10^{-12}$ & - & -\\
\\
{[}C$_4$dmim][NTf$_2$]\cite{Rodgers2009} & $4.4$ & $4.7 \times 10^{-10}$ & $7.5 \times 10^{-11}$ & $0.94 \pm 0.13 \times 10^{-3}$ & 0.4\\
\\
{[}N$_{6,2,2,2}$][NTf$_2$]\cite{Rodgers2009} & $4.5$ & $4.1 \times 10^{-10}$ & $7.3 \times 10^{-12}$ & $0.3 \times 10^{-3}$ & $0.36 \pm 0.03$\\
\hline
\end{tabular}
\end{center}
\caption{Literature values for Butler-Volmer fitting parameters for the reduction of oxygen in various ionic liquids}
\label{LIT VALUES}
\end{table}

\clearpage

\begin{table}
\begin{center}
\begin{tabular}{c c c}
Parameter & Butler-Volmer & Marcus-Hush \\
\hline
\\
$D_\mathrm{O_2}$ / m$^2$ s$^{-1}$ & $3.95 \times 10^{-10}$ &  $3.95 \times 10^{-10}$\\
\\
$c_\mathrm{O_2}$ / mM & 7.3 & 7.3 \\
\\
$k^0$ / cm s$^{-1}$ & 0.001 & 0.001 \\
\\
$E_\mathrm{f}^\minuso$ / V & $-1.47 \pm 0.01$ & $-1.55 \pm 0.01$ \\
\\
$\alpha$ & 0.30 & - \\
\\
$\lambda$ / eV & - & 0.5\\
\\
MSAD (average) / \% & 2.80 & 10.2\\
\hline
\end{tabular}
\end{center}
\caption{Simulation Results for the reduction of oxygen in [P$_{14,6,6,6}$][NTf$_2$] on a $\mu$Pt electrode.}
\label{P14666 PARAMETERS}
\end{table}

\clearpage

\providecommand*\mcitethebibliography{\thebibliography}
\csname @ifundefined\endcsname{endmcitethebibliography}
  {\let\endmcitethebibliography\endthebibliography}{}

\end{document}